\documentclass[pra,twocolumn,showpacs,preprintnumbers,amsmath,amssymb]{revtex4}

\usepackage{graphicx}
\usepackage{dcolumn}
\usepackage{bm}

\begin{document}

\preprint{}

\title{Spatial distribution of ions in a linear octopole radio-frequency 
ion trap in the space-charge limit}


\author{Takuya Majima}
\altaffiliation{Present address: Quantum Science and Engineering Center, Kyoto University, 
Gokasho, Uji, Kyoto 611-0011, Japan.}

\author{Gabriele Santambrogio}
\altaffiliation{Present address: Fritz-Haber-Institut der Max-Planck-Gesellschaft, 
Faradayweg 4-6, 14195 Berlin, Germany.}

\author{Christof Bartels}
\altaffiliation{Present address: Department of Dynamics at Surfaces, 
Max Planck Institute for Biophysical Chemistry, 
Am Fa\ss{}berg 11, D-37077 G\"{o}ttingen, Germany.}
\affiliation{East Tokyo Laboratory, Genesis Research Institute, Inc.,
717-86 Futamata, Ichikawa, Chiba 272-0001, Japan}

\author{Akira Terasaki}
\altaffiliation{Present address: Department of Chemistry, Kyushu University, 
6-10-1 Hakozaki, Higashi-ku, Fukuoka 812-8581, Japan}
\email{terasaki@chem.kyushu-univ.jp}

\author{Tamotsu Kondow}
\altaffiliation{Deceased.  May 25, 2009.}
\affiliation{Cluster Research Laboratory, Toyota Technological Institute \\
in East Tokyo Laboratory, Genesis Research Institute, Inc., 
717-86 Futamata, Ichikawa, Chiba 272-0001, Japan}

\author{Jan Meinen}

\author{Thomas Leisner}
\affiliation{Institute for Meteorology and Climate Research, 
Atmospheric Aerosol Research (IMK-AAF), 
Forschungszentrum Karlsruhe, 76021 Karlsruhe, Germany.}


\date{\today}

\begin{abstract}
We have explored the spatial distribution of an ion cloud trapped in a
linear octopole radio-frequency (rf) ion trap. The two-dimensional
distribution of the column density of stored Ag$_2^+$ was measured via
photofragment-ion yields as a function of 
the position of the incident laser beam over the transverse cross
section of the trap. The profile of the ion distribution was found to
be dependent on the number of loaded ions. Under high ion-loading
conditions with a significant space-charge effect, ions form a ring
profile with a maximum at the outer region of the trap,
whereas they are localized near the center axis region at low loading of
the ions. These results are explained quantitatively by a model
calculation based on equilibrium between the space-charge-induced
potential and the effective potential of the multipole rf field. The
maximum adiabaticity parameter $\eta_{\textrm{max}}$ is estimated to
be about 0.13 for the high ion-density condition in the present
octopole ion trap, which is lower than typical values reported for low
ion densities; this is probably due to additional instability caused
by the space charge.
\end{abstract}

\pacs{
37.10.Ty, 
41.90.+e, 
36.40.Mr  
}

\maketitle

\section{Introduction}

Linear multipole radio-frequency (rf) ion traps have become
increasingly important in atomic, molecular, and cluster
physics \cite{Reviews,Wester2009p154001}. In spectroscopic studies, for
example, these traps improve the resolution by enabling cooling with a
neutral buffer gas and improve the sensitivity by enabling
accumulation and thereby increasing the number
density \cite{Terasaki2007,Majima2008,Terasaki2009,Hirsch:2009p154029,Kostko:2007p012034,asmis:2002p1101,Goebbert:2009p7584,Rizzo:2009p481,Svendsen:2010p073107,Schlemmer:2002p2068,Asvany:2005p1219,Wolf:1995p4177}.
In gas-phase reactivity studies, multipole rf traps have proved ideal
in thermalizing all degrees of freedom of reagents and in providing
accurate knowledge of their
concentrations \cite{Paul:1995p373,Paul:1996p209,Socaciu:2003p10437}.
The success of these traps is due to the characteristics of the
time-averaged potential experienced by the trapped ions, the effective
potential $\Phi_\text{eff}$. The 
strength of the effective potential is proportional to $p^2$ for a
$2p$-pole trap, and its dependence on the distance $r$
from the central axis is proportional to $r^{2p-2}$ \cite{Gerlich1992}. 
For high values of $p$, traps are particularly deep and have a large
nearly field-free region around the axis, which guarantees
little rf heating \cite{Gerlich1992,Gerlich1995,Champenois2009,Gerlich2008}. The deeper a trap,
i.e.\ the larger the phase space acceptance, the larger the fraction
of ions that can be captured and accumulated. Additionally, buffer gas
cooling allows phase space compression, which, in turn, can
further increase the number of stored ions. Recently, the depth of a
22-pole trap was measured by analyzing the evaporation rates of
trapped ions from a thermodynamical point of view; this result was
employed to calculate effective trap depths for several $2p$-pole ion
traps as a function of the rf amplitude \cite{Mikosch2007,Mikosch2008}.

Taking advantage of the high ion density attainable in a multipole ion trap, 
we have recently been able to apply cavity ring-down spectroscopy
directly to the mass-selected ions stored in an octopole
trap \cite{Terasaki2007,Majima2008,Terasaki2009}. In the course of
these studies, the ions were found to be distributed not uniformly 
inside the trap, as it was first pointed out in
Ref.~\cite{Terasaki2007}. In fact, the ion density distribution is not
defined by the effective potential alone in a regime where space
charge effects become relevant. Understanding these
distributions in linear rf traps is extremely important 
(i) for the optimization of spectroscopic methods that
profit from the maximal overlap of laser light with the ion
clouds inside the trap, 
(ii) for the measurement of absolute absorption cross sections, 
(iii) to accurately estimate the thermalization temperature in reactivity studies, 
and (iv) for optimization and interpretation of experiments where the ion trap is
used as a pick-up cell \cite{Bierau2010p133402}. 

The density distribution of a non-neutral plasma in a Penning trap has
been extensively studied by Dubin and O'Neil \cite{Dubin1999}.
For a Paul trap, i.e., a quadrupole rf field, 
profiles of an ion cloud were studied for atomic metal ions and
organic molecular ions by monitoring fluorescence and
photodissociation of stored ions, respectively  
\cite{Siemers1988,Hemberger1992,Cleven1996}.
These profiles have a Gaussian shape with the maximum ion density at
the center of the trap. More recently, even single ions have been
observed by fluorescence imaging of Coulomb crystals formed in a
linear Paul trap at temperatures around 10~mK \cite{Mortensen2006}. For
multipole ion traps, on the other hand, only a few measurements have
been reported. Walz et al.\ measured a radial distribution of ions in
a three-dimensional hyperbolic octopole ion trap by monitoring
fluorescence intensity from stored Ba$^+$ ions \cite{Walz1994}. They
found that Coulomb repulsion between stored ions resulted in two
separate ion peaks. Wester and coworkers reported a radial
distribution of column densities of OH$^-$ ions in a 22-pole ion trap
from photodetachment-rate measurements \cite{Trippel2006}.  The
distribution showed a rather uniform profile and was explained by a
model calculation neglecting Coulomb-repulsion effects; this model was
applicable to the measurement performed at a low ion density (less
than $10^3$ stored ions). They extended the measurement to
two-dimensional tomography of the column density \cite{Otto2009}.  A
Coulomb crystal of laser-cooled Ca$^+$ ions has recently been formed
in a linear octopole ion trap by Okada et al.; the ions were observed
by fluorescence imaging for storage of up to $10^4$ ions, which is
still in the low-density regime \cite{Okada2007,Okada2009}. 

In this paper, we report on the measured ion-density profile in a
linear octopole ion trap at a high-density regime, where the space
charge plays a significant role. Silver dimer cations, Ag$_2^+$, are
detected via photofragmentation and the 
two-dimensional distributions of column densities are measured as a
function of the number of loaded ions and as a function of the
amplitude of the rf field. Up to about $10^9$ ions are loaded into the
trap, which is the space-charge-limit condition for the present ion
trap geometry. 
The distribution profiles are compared with a model calculation, 
which takes the balance between the trapping force due to the rf field 
and the Coulomb repulsion among the stored ions into account.

\section{Experimental Procedures}\label{Expt}

\begin{figure}[b]
  \includegraphics{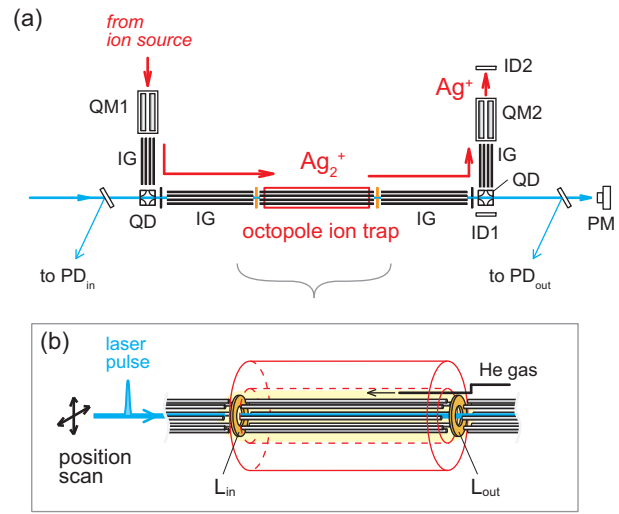} 
  \caption{\label{Setup}
  A schematic diagram of the experimental setup.
  QM1 and QM2, quadrupole mass filters;
  IGs, octopole ion guides;
  QDs, quadrupole deflectors; 
  PD$_\text{in}$ and PD$_\text{out}$, photodiodes;
  PM: a power meter;
  L$_\text{in}$ and L$_\text{out}$, entrance and exit electrodes of the ion trap;
  ID1 and ID2, ion current detectors.
  }
\end{figure}

A schematic diagram of the experimental setup is shown in Fig.~\ref{Setup};
a detailed description can be found elsewhere \cite{Terasaki2007}.
Ag$_2^+$ ions were selected by a quadrupole mass filter (labeled QM1,
model MAX-4000 by Extrel CMS) from the distribution generated by a 
magnetron-sputter cluster-ion source. A typical current
of the Ag$_2^+$ ion beam after the mass filter was 200~pA, with QM1
operated at relatively low mass resolution ($m/\Delta m \sim
20$). The selected Ag$_2^+$ ions were transported by octopole ion
guides (IGs) and quadrupole deflectors (QDs) to a
linear octopole ion trap.

The ion trap has a length of 40~cm and an inner diameter of
1.1~cm. 
The trap is filled with He buffer gas to thermalize all degrees of
freedom of the ions. All data presented here were measured at room
temperature. The rf potentials at about 3~MHz applied to the
octopole's rods were provided by a homemade rf
generator \cite{Jones2000}. The rf amplitude $V_\text{rf}$ was varied 
in the range 95--300~V, while the offset dc potential was kept at $-12$ V.
The ions were confined in the longitudinal direction using potentials
generated by an entrance (L$_\text{in}$) and an exit (L$_\text{out}$) electrode.
The potential of L$_\text{in}$ was switched between $-8$ and $+5$~V for loading 
and trapping, respectively. L$_\text{out}$ was held at
$+5$~V during loading and trapping, and was switched to $-12$~V for
extraction of trapped ions. 
The number $N_0$ of ions stored in the trap was measured via
monitoring the ion current at the detector (ID1), taking into account
the transmission probabilities of IG and QD. These were estimated to
be unity and 0.9, respectively. By controlling
the loading time, the ion current, and the rf amplitude, $N_0$ was
varied from $4.0 \times 10^7$ up to $1.2 \times 10^9$; the latter is
the maximum ion loading permitted by the present ion trap.

The column density of the ions was measured by recording the
photofragmentation yield introduced by UV laser pulses at 415~nm, 
near the peak absorption of Ag$_2^+$ \cite{Egashira2011}. 
These were generated by using an optical parametric oscillator (MOPO-HF, Spectra
Physics) operated at a repetition rate of 10 Hz, which was attenuated to 
about 20~$\mu$J/pulse. 
With a collimated beam of 2~mm diameter, the intensity was sufficiently low
that the photofragment yield depended linearly on the laser pulse
energy. The laser beam was aligned carefully to
be parallel to the axis of the ion trap. To map the ion density
distribution, the laser position was displaced both vertically and
horizontally with an interval of 0.5~mm. The laser pulse intensity was
monitored during the measurement by the photodiodes (PD$_\text{in}$
and PD$_\text{out}$) 
located before and after the ion trap, outside the vacuum chamber. The
signal intensities of the PDs were calibrated by a power meter (PM)
placed downstream. The number of photons of the injected laser pulses,
$N_p$, was determined from the signal intensities of PD$_\textrm{in}$
measured shot by shot. The intensity ratio between the two PDs was used
to confirm that the laser beam was not clipped while scanning its
position. 
The number of photofragment ions, $N_f$, are extracted, mass analyzed
using a quadrupole mass filter (QM2) and detected at ID2. The
transmission probability of QM2 was estimated to be 0.4.

The measurement was performed following this procedure: First, Ag$_2^+$
ions were loaded into the ion trap for a duration between 0.3 and
2.0~s depending on the number of ions to be stored. Second, the stored 
ions were thermalized by collisions with the He buffer gas at room
temperature for 0.5~s. Third, the ions were irradiated with the laser
pulse for 1~s, i.e., 10 shots.  Finally, the ions were extracted from
the trap and the yield of  Ag$^+$ photofragments was recorded. The
above measurement was repeated five times at each laser position. 

An absolute value of the local number density of ions,
$n(\bm{r})$, was derived from the relationship:
\begin{equation}
	n(\bm{r})= \frac{f(\bm{r})}{\sigma \, L},
\end{equation}
where $\sigma$ is the photodissociation cross section, $L$ is the
length of the ion trap, and $f(\bm{r})$ is the number of fragments
ions per photon ($N_f / N_p$).  We measured the quantity $f$ as a
function of the laser position $\bm{r}$. Because the trap is
40~cm long, we assume a uniform distribution of ions along the
longitudinal direction of the linear ion trap. The cross section  
$\sigma$ was determined from the normalization condition:
\begin{equation} \label{Norm}
N_0 = L \int n(\bm{r}) d\bm{r} = \frac{1}{\sigma}\int f(\bm{r}) d\bm{r}.
\end{equation}


\section{Results}

\begin{figure}[b]
  \includegraphics{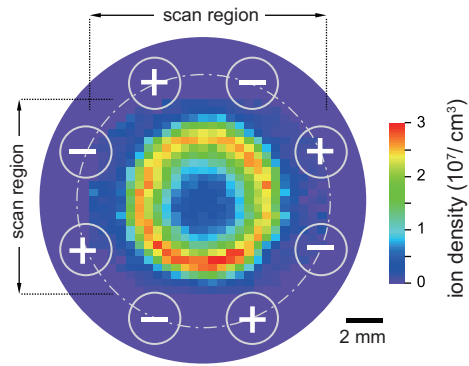}
  \caption{\label{2Dscan_8config}
  Two-dimensional distribution of the Ag$_2^+$ ion density.
  The total number $N_0$ of stored ions was about $5.5 \times 10^8$.
  The eight circles represent the pole electrodes of the ion trap.}
\end{figure}

Figure~\ref{2Dscan_8config} shows the result of a two-dimensional scan of
an ion cloud containing $5.5 \times 10^8$ Ag$_2^+$ ions. A
major part of the trapped ions are found in the outer region rather
than at the center of the ion trap. The small variation of the densities
depending on the azimuthal angle might be caused  by imperfect
configuration of the poles and/or other neighboring
electrodes \cite{Otto2009}.  In the following discussion, we assume
cylindrically symmetric distributions and will analyze the radial
distributions obtained by one-dimensional scans along the horizontal
axis. These data are then analyzed using an adiabatic
approximation, which provides a rotationally symmetric shape for the
effective potential \cite{Gerlich1992}.

\begin{figure}[b]
  \includegraphics{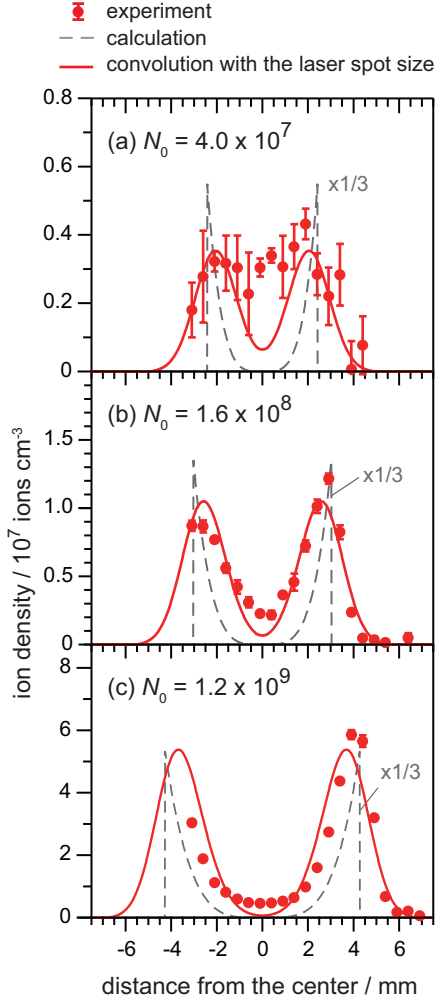}
  \caption{\label{Ndep}
  Radial distributions of the ion density obtained by one-dimensional scans 
on the horizontal axis for different amounts of loaded ions.
The amplitude of the rf field $V_\text{rf}$ was 200~V.
Error bars show statistical standard errors of the measurements. 
Dashed lines show calculated distribution based on an adiabatic
approximation according to Eq.~\eqref{rho_T0eff} using the measured
values for $N_0$.
Upon convolution the calculated distribution with a
Gaussian function of 2~mm FWHM representing the 
laser beam profile, we obtained the solid lines which should be
compared with the experimental data. Thermal effects are not
considered in these calculations.} 
\end{figure}

The radial distributions of the ion density are shown in Fig.~\ref{Ndep} 
for three different amounts of loaded ions, along with model
calculations discussed in the following section. For these
measurements $V_\text{rf}$ was held constant at 200~V, the optimal
value.  With $N_0= 4.0 \times 10^7$, the ion distribution is found to
be concentrated around the center of the trap (Fig.~\ref{Ndep}(a)).
As $N_0$ is increased, the ion density in the central region increases
only slightly while most of the ions are found in the outer region. At
the maximum loading condition of $N_0= 1.2 \times 10^9$, the ion
density is peaked around $r=4$~mm (Fig.~\ref{Ndep}(c)). 
The fact that the density distribution is confined to the center of the
trap when the ion number is low suggests that the ions are well thermalized 
by collisions with the He buffer gas. Therefore, we interpret the ring
profile of the distribution measured for the highest ion density,
$N_0= 1.2 \times 10^9$, as the effect of Coulomb repulsion between
the trapped ions. 

\begin{figure}[b]
  \includegraphics{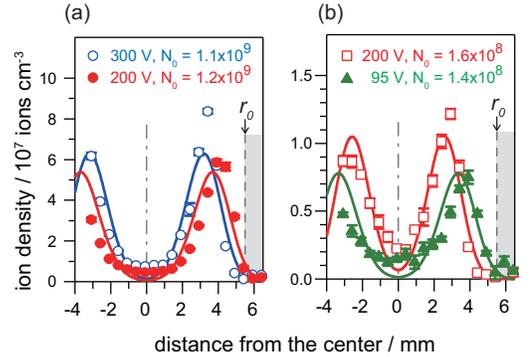}
  \caption{\label{RFdep}  
Experimental (symbols) and calculated (line) radial distribution profiles 
of the ion density for various rf amplitude $V_\text{rf}$.  
(a) The trap is fully loaded with $N_0=1.1 \times10^9$ and 
$1.2 \times 10^9$ ions at $V_\text{rf}=300$ and 200~V, respectively. 
(b) The trap is fully loaded with $N_0=1.4 \times10^8$ ions at $V_\text{rf}=95$~V 
and partially loaded with $1.6 \times 10^8$ ions at $V_\text{rf}=200$~V 
(the latter are the same data as in Fig.~\ref{Ndep}(b)). 
} 
\end{figure}

The radial distribution profiles are shown in Fig.~\ref{RFdep} for
three values of the rf amplitude $V_\text{rf}$. Ions were loaded
until saturation for each $V_\text{rf}$; the number of ions was
measured to be $1.4 \times 10^8$, $1.2 \times 10^9$, and $1.1 \times 10^9$ for
$V_\text{rf}=95$, 200, and 300~V, respectively. The number of ions
increased by a factor of 8.6 when $V_\text{rf}$ was changed from 95 to
200~V, while it slightly decreased when $V_\text{rf}$ was changed from
200 to 300~V. Although the total number of 
ions varied only slightly between $V_\text{rf}=200$ and 300~V, the distributions
exhibited clearly different profiles; with the higher rf amplitude the
density peak became sharper and was shifted by 0.5~mm toward the
center of the trap. 
In Fig.~\ref{RFdep} (b), the measurement at $V_\text{rf}=200$~V shown in 
Fig.~\ref{Ndep} (b), obtained without filling the trap, is superimposed 
for comparison. 
We note that, although the total number of ions in Fig.~\ref{Ndep} (b) and 
that in the trap filled at $V_\text{rf}=95$~V are approximately the same, 
the distributions are different.

In the above analyses of the absolute ion densities, the
photodissociation cross section, $\sigma$, was evaluated by
Eq.~\eqref{Norm} for each measurement. It was found to be
$\sigma$ = $(5 \pm 1) \times 10^{-17}$~cm$^2$ 
at 415~nm for the present 300-K ion trap. 
The uncertainty represents a statistical error of the measurement; 
we mention that this evaluation may have an additional systematic error 
due to uncertainties in the estimation of the ion transmittance through 
the ion optics (see Section~\ref{Expt}) and in the measurement of the laser pulse energy.
Note that, as the spectral profile of photoabsorption is dependent on 
the temperature of the ions, the cross section at a given wavelength 
changes accordingly \cite{Egashira2011}.

\section{Model calculation}

To explain the radial distributions of the ion density dependent on
the trapping conditions, we have performed a model calculation based
on an adiabatic approximation \cite{Gerlich1992}.  Here, effects of the
static potential produced by L$_\text{in}$ and L$_\text{out}$ are
neglected because they are much smaller than the magnitude of the
effective potential for the long trap we have used (40~cm). Thus
we assume a cylindrical symmetry along the trap 
axis. The transverse motion of the ions in a multipole ion trap is
described as a motion in an effective electric potential 
\footnote{Our
  notation differs from the one used by Gerlich~\cite{Gerlich1992}, in
  that we divide the effective mechanical potential by the charge $q$
  in order to obtain an \emph{effective electric potential}.}
expressed by 
\begin{equation}\label{Phi_eff}
  \Phi_\text{eff}(r) = 
  \frac{p^2}{4} \frac{q\, V_\text{rf}^2}{m\, \Omega^2\, r_0^2} \bigg(\frac{r}{r_0}\bigg)^{2p-2},
\end{equation}
where $q$, $m$, $2p$, $\Omega/2\pi$, and $r_0$ denote the ion charge,
the ion mass, the number of poles, the frequency of the rf field, and
the inscribed radius of the ion trap, respectively \cite{Gerlich1992}. 
In general, the equilibrium ion density $n(r)$ in such a potential at
a given temperature $T$, is described by 
\begin{equation}\label{rho_general}
n(r) = n_0 \exp{\bigg[-\frac{q}{k_B\, T}\bigg(\Phi_\text{eff}(r) + \Phi_\text{sc}(r)\bigg)\bigg]},
\end{equation}
where $n_0$ and $k_B$ are a normalization and Boltzmann's constants,
respectively, and $\Phi_\text{sc}(r)$ is the electric potential due to
the space charge of the ion cloud \cite{Walz1994,Champenois2009}.
In the general case, Eq.~\eqref{rho_general} is non-linear
because $\Phi_\text{sc}(r)$ is 
related to the local ion density by Poisson's equation,
\begin{equation}\label{poisson}
\nabla^2 \Phi_\text{sc}(r)=-\frac{ q \; n(r)}{\varepsilon_0}.
\end{equation}
However, if one assumes the space charge effect to be negligible,
Eq.~\eqref{rho_general} simplifies to 
\begin{equation}
n(r) = n_0 \exp{\bigg[-\frac{q}{k_B\, T}\Phi_\text{eff}(r)\bigg]}.
\end{equation}
This approximation was applied to explain an ion distribution in a
22-pole trap containing about 10$^3$ ions \cite{Trippel2006}. In
the present study, we have about 5 to 6 orders of magnitude more ions so
that we load the trap until saturation. Therefore we cannot
neglect the space charge effect, and as a result we rearrange
Eq.~\eqref{rho_general} as 
\begin{equation}
-\Phi_\text{sc}(r)=\frac{k_B\,T}{q}\ln\frac{n(r)}{n_0}+\Phi_\text{eff}(r).
\end{equation}
At zero temperature the space charge due to the ion
distribution must exactly counterbalance the effect of the external
potential $\Phi_\text{eff}$. That is to say, neglecting the energy due
to the thermal motion of the ions allows one to recast
Eq.~\eqref{rho_general} as~\cite{Dehmelt1967}
\begin{equation}\label{eq:T0}
\Phi_\text{eff}(r) +\Phi_\text{sc}(r) = 0.
\end{equation}
Equations~\eqref{poisson} and \eqref{eq:T0} thus lead to 
\begin{equation}\label{poisson_PHIeff}
n(r) = \frac{\varepsilon_0}{q}\;\nabla^2 \Phi_\text{eff}(r), 
\end{equation}
which determines the radial profile of the ion density.  
We will see later on that this approximation is not entirely fulfilled,
but it is nevertheless sufficiently good to explain most of our experimental
data. 
An alternative approach to the general derivation of the charge
distribution in a cylindrically symmetrical external potential is
described in the Appendix. 
Using Eq.~\eqref{Phi_eff} for the effective potential, 
we obtain the following radial distribution from Eq.~\eqref{poisson_PHIeff}:  
\begin{equation}\label{rho_T0eff}
n(r) = p^2 (p-1)^2 \frac{\varepsilon_0 V_\text{rf}^2}{m \Omega^2 r_0^4}\bigg(\frac{r}{r_0}\bigg)^{2p-4}.
\end{equation}
At $T=0$ we expect a sharp cut in the distribution at $r_\text{max}$,
which is determined by the number $N_0$ of stored ions. In an octopole
ion trap ($p = 4$), the ion density is proportional to $r^4$. Note
that instead, in the case of a quadrupole ($p=2$), 
Eq.~\eqref{rho_T0eff} predicts a constant ion distribution.
In order to compare these predictions with the experimental data, 
the distribution $n(r)$ was evaluated in the interval $[0,r_\text{max}]$ 
and then convoluted with a Gaussian function, 2~mm FWHM, 
representing the laser-beam diameter. 
The results of the calculations are shown in Figs.~\ref{Ndep} and~\ref{RFdep} 
together with the experimental data. 
The maximal radius $r_\text{max}$ is such that the integral
of $n(r)$ over a 40-cm long cylinder of radius $r_\text{max}$
corresponds to the measured $N_0$. 

\begin{figure}[h!]
  \includegraphics{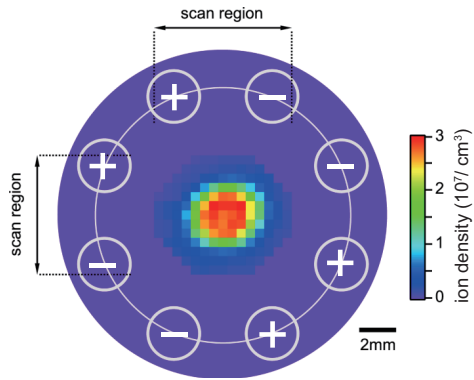}
  \caption{\label{2Dscan_4config}
A two-dimensional distribution of ion densities measured for Ag$_2^+$ ions 
in a quasi-quadrupole ion trap, which should be compared with
Fig.~\ref{2Dscan_8config}. In this measurement the neighboring poles
were wired pairwise together to form four pairs of electrodes. The
total number of ions stored was about $1.3 \times 10^8$ 
in this particular condition with $V_\text{rf}$ = 360~V. The eight
circles represent the pole electrodes of the ion trap, along with the
polarity of the RF voltage applied to each electrodes.} 
\end{figure}

The standard way to create multipole potentials is by
approximating the hyperbolic equipotential surfaces with cylindrical
electrodes. In general, it is always possible to approximate a lower
order multipole with the electrodes configuration designed for a
higher order one. Thus, we approximated a quadrupole by wiring the
neighboring poles of our octopole trap together to form four pairs
of electrodes. The ion distribution measured for this 
quasi-quadrupole ion trap is shown in Fig.~\ref{2Dscan_4config}.
The ions are concentrated around the axis with a rather flat density
profile as predicted by the present model calculation. As the surface of
the eight rods cannot perfectly follow the equipotential lines of a
quadrupole potential, this quadrupole trap is not perfect
and provides a smaller trapping volume than an ordinary four
rods configuration.

\section{Discussion}

The distributions calculated by using Eq.~\eqref{rho_T0eff}
are compared with experimental results in Fig.~\ref{Ndep}. Note that
three curves calculated for different $N_0$ have the same profile
except for $r_\text{max}$,  which was determined to be 4.3, 3.0, and
2.4~mm for $N_0=1.2 \times 10^9$, $1.6 \times 10^8$, and $4.0 \times
10^7$, respectively. The calculated curves reproduce well the overall
features of the experimental results, particularly for larger $N_0$.
This agreement indicates that the ring profiles of the ion
distribution are mainly governed by the space charge effect, which
forces the stored ions toward the outer
region. Note that the largest discrepancies between the model and the
experiment are found near the center of the trap and for
a low ion density. This is due to the $T=0$ approximation, which is a
poor approximation where $\Phi_\text{eff}$ and $\Phi_\text{sc}$
are comparable to the energy of the thermal motion, i.e., at small
$r$ and/or small $N_0$. The room temperature, $k_B T=26$~meV, is equal to
$\Phi_\text{eff}(r)$ at $r=2.2$~mm for $V_\text{rf}=200$~V. 

Figures~\ref{2Dscan_8config} and~\ref{2Dscan_4config} show that the
maximal ion densities in an octopole and in a quadrupole are
similar, but the latter provides the maximum  
around the center of the trap, whereas in the former the maximal
density is found on a ring at larger radius. Since it is easier to
overlap a laser beam to an ion cloud that is concentrated around the
center of the trap than to a ring, a quadrupole is probably the most
favorable rf linear trap for laser spectroscopy. However, since the
phase space acceptance of a quadrupole is lower than that of every
other higher order linear rf trap, it is possible that many ions
injected into a quadrupole are lost. This makes the loading time of a
quadrupole longer than for a higher oder multipole. In order to
optimize the loading time \emph{and} the laser beam overlap with the
ions, one can imagine using a high-order multipole trap for loading and
thermalizing the ions and then switching it to a quadrupole for
spectroscopy. This can be done by controlling the potential on each
electrode independently.

According to Eq.~\eqref{rho_T0eff}, the maximum radius
$r_\text{max}$ within which ions are stably stored is determined by
the total number of ions and by $V_\text{rf}$---for a given trap
geometry and a given $\Omega$. For full traps, however, two other
factors reduce the value of $r_\text{max}$. One is the physical geometry of the
traps: according to Gerlich \cite{Gerlich1992} a typical geometric
limit $r_\text{geom}$ is $0.8\,r_0$ due to the space necessary for
micro-motion wiggling. The other is the breakdown of the adiabatic
approximation due to large rf amplitude. The adiabaticity is
quantified by means of a dimensionless adiabaticity parameter
$\eta(r)$, as described in Refs. \cite{Gerlich1992} 
and \cite{Teloy1974}:    
\begin{equation}
\eta(r) = 2p (p-1) \frac{qV_\text{rf}}{m \Omega^2 r_0^2}\bigg(\frac{r}{r_0}\bigg)^{p-2}.
\end{equation}
There is a maximum value of $\eta=\eta_\text{max}$ for which the rf
heating makes the motion of the ions unstable. Since $\eta(r)$
increases with $r$, there is a critical radius $r_\text{c}$ at which 
$\eta(r_\text{c}) = \eta_\text{max}$. 
Whereas $r_\text{c}$ decreases with increasing $V_{\textrm{rf}}$,
$r_\text{geom}$ is independent of $V_{\textrm{rf}}$. Therefore, 
$r_\text{max}$ is equal to $r_\text{geom}$ at a sufficiently low
$V_\text{rf}$. As $V_\text{rf}$ grows, the number of stored ions
increases according to Eq.~\eqref{rho_T0eff} until $r_\text{c}$ is
reduced to $r_\text{geom}$. A further increase of $V_\text{rf}$ causes
a loss of the ions because $r_\text{max}$ (now equal to $r_\text{c}$)
is reduced. We summarize these two effects as
$r_\text{max}=\text{min}[r_\text{c},r_\text{geom}]$. This behavior has
been observed by Mikosch et 
al. \cite{Mikosch2008} in their measurements of the trapping potential
depth.

The reduction of the maximum radius $r_\text{max}$ observed when
$V_\text{rf}$ is changed from 200 to 300~V as shown in 
Fig.~\ref{RFdep} is attributed to the decrease of $r_\text{c}$.
For the $V_\text{rf}=300$~V case, $\eta_\text{max}$ is determined to
be 0.13, where the value $r_\text{max} = 3.7$~mm is obtained from
Eq.~\eqref{rho_T0eff} and from the measured $N_0$. We note that this
value for $r_\text{max}$ is in good agreement with the data shown in
Fig.~\ref{RFdep}. Other groups found $\eta_\text{max} =0.36 \pm 0.02$
for a 22-pole trap \cite{Mikosch2008} and $\eta_\text{max} \sim 0.2$
for a 3D octopole trap \cite{Walz1994}. In addition, Gerlich suggested
$\eta_{\textrm{max}}= 0.3$ from numerical
simulation \cite{Gerlich1992}.  Our value is much lower than the
one for the 22-pole trap and the calculated one. The 22-pole
experiment was done with less than
10$^3$ ions. Gerlich performed the simulation for a single ion free from
perturbation by other ions. In contrast, the present experiment was
performed under a strong space-charge effect. The reduction of
$\eta_\text{max}$ indicates that the ion--ion interaction introduced
an additional source of instability.

On the premise that the same $\eta_\text{max}$ = 0.13 be applied for 
the lower $V_\text{rf}$, the values of $r_\text{c}$ were calculated to
be 4.4 and 6.4~mm for $V_\text{rf}=200$ and 95~V, respectively.
Clearly, $r_\text{max}$ is limited by $r_\text{geom}$ at $V_\text{rf}=95$~V.
At $V_\text{rf}=200$~V we were able to store the maximum number of ions and
therefore $r_\text{max}=r_\text{c}=r_\text{geom}$; the $r_\text{c}$ value of 4.4~mm 
is in good agreement with the $r_\text{max}$ value extracted from Fig.~\ref{RFdep}.
The value of $r_\text{geom}/r_0 = 4.4/5.5=0.8$ is consistent with 
Gerlich's estimation of $r_\text{geom}/r_0 \leq 0.8$. 

The values of $r_\text{geom}$ and $\eta_\text{max}$ are
specific of this trap geometry and possibly valid only in the high
density limit, but otherwise independent of any other experimental
parameter. We have extracted these values using the measured number of
ions $N_0$ together with Eq.~\eqref{rho_T0eff}, which describes
the ion distribution in the high density limit. As the measurement of
absolute numbers is always challenging, it is interesting
to consider the inverse problem: namely, the determination of the
absolute number of ions in the trap, $N_0$, based on a relative
measurement of the ion density and on the knowledge of the
characteristics of the trap. This is only possible
when the densities are high enough for the interaction between
ions to become relevant in shaping the ion distribution. Then, one
can turn a relative measurement of the ion distribution into a
measurement of the interaction strength between ions and, thus,
into a measurement of the absolute number of ions. 
Close to the space charge limit, one obtains $N_0$ directly by 
integration of Eq.~\eqref{rho_T0eff} between $0$ and $r^0_\text{max}$, 
where $r^0_\text{max}$ is the maximum of the ion distribution, 
which is determined experimentally.

\section{Summary}

We studied the radial distributions of ions 
stored in a linear octopole ion trap near the space charge limit by
monitoring photofragmentation yields as a function of the laser 
position. For the highest densities, we observed that
the ion distribution has a ring profile. We showed that this is a
typical feature of a multipole ion trap. The quadrupole potential,
however, is an exception in the family of the linear multipole rf
traps as it induces a uniform ion distribution even when the space
charge limit is reached. These observations are predicted and
explained by a simple model based on equilibrium between the effective 
potential and that produced by the charge of the ions. The only
approximation we used is that the energy related to the thermal motion
of the ions inside the trap is negligible in comparison with the
effective potential generated by the multipole. This approximation is
fulfilled except for the regions where the effective potential is very
flat, which, however, contain very few ions when the trap is full.

The maximum adiabaticity parameter, $\eta_\text{max}$, was estimated
to be 0.13, when the trap is full. This value is lower than those
found in other studies under low ion-density conditions. We
tentatively attribute the reduced value of $\eta_\text{max}$ to an
additional source of instability induced by repulsive forces among
the stored ions. 

Understanding the space charge effects allows to
extact the absolute number of ions in the trap and 
their absolute density based on the relative distribution. This
provides a way of measuring absolute numbers without knowledge of
absorption cross sections nor detector efficiencies.

\begin{acknowledgments}
The present study was supported by the Special Cluster Research Project 
of Genesis Research Institute, Inc.
\end{acknowledgments}

\appendix*
\section{Charge distribution in a cylindrically symmetrical external potential}

The derivation of Eq.~\eqref{poisson_PHIeff} is presented in an alternative way 
for a general cylindrically symmetrical external potential.  
The charge distribution is derived under the assumption that it is translational
invariant along the $z$-axis, where it extends from negative to
positive infinity. Thus, all extensive quantities as charge or energy represent 
a ``per unit length'' value in the following.  

Assume that a certain amount of charge per unit length, $Q_0$, is
allowed to distribute freely in a cylindrically symmetrical external 
potential $\Phi_\text{ext}$.
  In cylindrical coordinates $r$, $z$, and $\theta$,
\begin{equation}
\Phi_\text{ext}(r,z,\theta)=\Phi_\text{ext}(r).
\end{equation}
  The charge will distribute radially with density $\rho(r)$ as to
minimize the total electrostatic energy $E$ per unit length of $Q_0$
in $\Phi_\text{ext}$.
  The density, $\rho(r)$, is to be calculated as follows for any given 
external potential $\Phi_\text{ext}(r)$.

It is convenient to introduce the cumulative charge $Q(r)$, which is
the amount of charge per unit length within a cylinder of radius $r$
around the $z$-axis.
  $Q(r)$ and $\rho(r)$ are related by
\begin{equation}\label{eq:q-rho}
\rho(r)=\frac{1}{2\pi r}Q'(r),
\end{equation}
where $Q'(r)$ is the derivative of $Q(r)$ with respect to $r$ and
$Q(r)$ is defined for $0\leq r \leq \infty$, its boundary values are 
$Q(0)=0$ and $Q(\infty)=Q_0$.
  The total energy $E$ per unit length of the charge distribution is
composed of the internal energy $E_\text{i}$ of $\rho(r)$ 
(i.e., the electrostatic energy resulting from the repulsion between the volume
elements of $\rho(r)$) 
and the external energy, $E_\text{e}$, of the charge distribution $\rho(r)$ 
in the external potential.
  The latter is found by volume-integration of the product $\rho(r) \Phi_\text{ext}(r)$ 
and can be expressed with the help of Eq.~(\ref{eq:q-rho}) as a one dimensional
integral along the $r$-coordinate:
\begin{equation}\label{eq:E-ext}
E_\text{e}=\int_A \rho(r)\Phi_\text{ext}(r)\,dA=\int_0^\infty Q'(r)\Phi_\text{ext}(r)\, dr,
\end{equation}
where $A$ represents the volume in the $r$--$\theta$ plane.
  To access $E_\text{i}(r)$ it is convenient to derive the strength of
the radial internal field $\bm{E}$ from the first Maxwell equation
applied over the surface of an infinite cylinder of radius $r$ along
the $z$-axis:
\begin{equation}
\bm{E}=\frac{Q(r)}{2\pi \varepsilon_0 r}\bm{e}_r.
\end{equation}
  From classical electrostatics it follows that $E_\text{i}$ is given by
\begin{equation}\label{eq:E-int}
E_\text{i}=\frac{\varepsilon_0}{2}\int_A\bm{E}\cdot\bm{E}\,
dA=\int_0^\infty \frac{Q^2(r)}{4\pi\varepsilon_0 r}dr.
\end{equation}
  Combining Eqs.~\eqref{eq:E-ext} and~\eqref{eq:E-int}, the total energy
can be written as a single integral over $r$
\begin{equation}\label{eq:E-tot}
E=\int_0^\infty\bigg\{ Q'(r)
\Phi_\text{ext}(r)+\frac{Q^2(r)}{4\pi\varepsilon_0 r}\bigg\}\,dr.
\end{equation}
  Equation~\eqref{eq:E-tot} is already in the canonical form for the
calculus of variations.
  The distribution $Q(r)$ that minimizes $E$ under the given boundary 
conditions can be found by solving the Euler-Lagrange equation
\begin{equation}\label{eq:var}
\frac{\partial}{\partial r}\frac{\partial L}{\partial
  Q'}-\frac{\partial L}{\partial Q}=0,
\end{equation}
where $L=\big\{Q'(r)\Phi_\text{ext}(r)+\frac{Q^2(r)}{4\pi \epsilon_0 r}\big\}$, 
the integrand of Eq.~(\ref{eq:var}), is the Lagrangian associated with 
the variational problem.
  Equation~\eqref{eq:var} leads to 
\begin{equation}\label{eq:Q}
Q(r)=2\pi\varepsilon_0 r\Phi_\text{ext}'(r),
\end{equation}
which can be used to finally derive $\rho(r)$ with the help of
Eq.~(\ref{eq:q-rho}):
\begin{equation}\label{eq:rho}
\frac{\rho(r)}{\varepsilon_0}= \bigg\{
\frac{\Phi_\text{ext}'(r)}{r}+\Phi_\text{ext}''(r)\bigg\}=\nabla^2\Phi_\text{ext}(r) 
\end{equation}
which is equivalent to Eq.~\eqref{poisson_PHIeff} and is applicable to
calculate $\rho(r)$ for any given cylindrically symmetric potential
$\Phi_\text{ext}(r)$.

\bibliography{bib_v2}

\end{document}